\documentclass[sigconf]{acmart}
\usepackage{algorithm}
\usepackage{algorithmic}
\usepackage{graphicx}
\usepackage{subcaption}
\usepackage{url}
\usepackage{placeins} 
\usepackage{float} 
\renewcommand\footnotetextcopyrightpermission[1]{} 
\AtBeginDocument{%
  \providecommand\BibTeX{{%
    \normalfont B\kern-0.5em{\scshape i\kern-0.25em b}\kern-0.8em\TeX}}}

\acmConference[preprint]{on}{arXiv}
\graphicspath{{./images/}}

\begin{document}

\title{Aligning LLMs with Human Instructions and Stock Market Feedback in Financial Sentiment Analysis}

\author{Zijie Zhao}
\affiliation{%
 \institution{Massachusetts Institute of Technology}
 \city{Cambridge}
 \state{MA}
 \country{USA}}

\email{zijiezha@mit.edu}
\author{Roy E. Welsch}
\affiliation{%
 \institution{Massachusetts Institute of Technology}
 \city{Cambridge}
 \state{MA}
 \country{USA}}
\email{rwelsch@mit.edu}

\renewcommand{\shortauthors}{Zhao and Welsch, et al.}

\begin{abstract}
Financial sentiment analysis is crucial for trading and investment decision-making. This study introduces an adaptive retrieval augmented framework for Large Language Models (LLMs) that aligns with human instructions through Instruction Tuning and incorporates market feedback to dynamically adjust weights across various knowledge sources within the Retrieval-Augmented Generation (RAG) module. Building upon foundational models like LLaMA 2, we fine-tune a series of LLMs ranging from 7B to 70B in size, enriched with Instruction Tuning and RAG, and further optimized through direct feedback and Reinforcement Learning (RL)-based refinement methods applied to the source weights of RAG. Through extensive evaluation, we demonstrate that the sentiment outputs from our LLMs more accurately mirror the intrinsic sentiment of textual data, showcasing a 1\% to 6\% boost in accuracy and F1 score over existing state-of-the-art models and leading conversational AI systems. Moreover, the sentiments extracted are more indicative of the directions in stock price movements. On top of that, we successfully construct portfolios that yield a 3.61\% higher Sharpe ratio compared to the S\&P 500 baseline in bullish markets. These portfolios also demonstrate resilience in bearish markets, with a 5x reduction in return losses compared to those typically experienced by the S\&P 500.

\end{abstract}




\maketitle

\section{Introduction}
\label{sec:intro}

\begin{figure*}[!ht]
\centering
\includegraphics[width=0.8\textwidth]{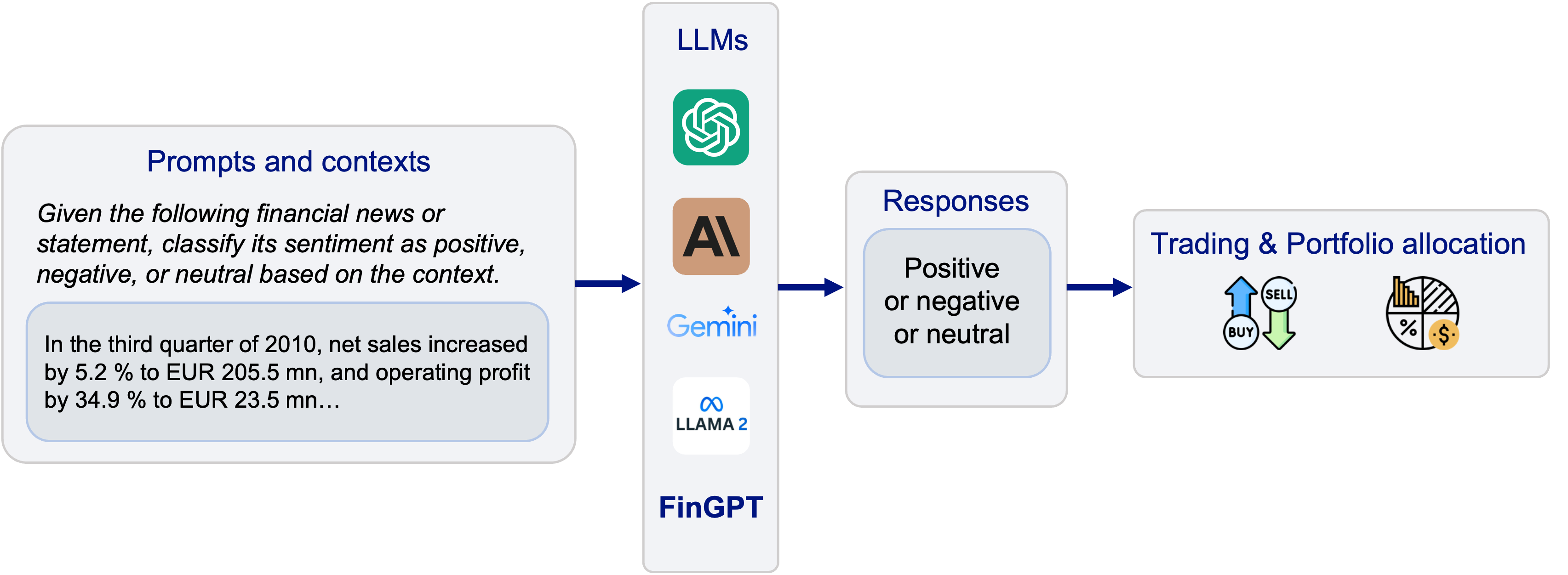}
\caption{Workflow of financial sentiment analysis using LLMs.}
\label{fig:overview}
\end{figure*}

Financial sentiment analysis is essential in interpreting investor sentiment derived from various sources such as financial articles, news, and social media. This analysis plays a pivotal role in economic forecasting, corporate finance decisions, and particularly in understanding and predicting market trends. By leveraging sentiment outputs—categorized as positive, negative, or neutral—researchers and practitioners can formulate effective trading strategies and optimize portfolio allocation. Recent advancements in Large Language Models (LLMs) have demonstrated their efficacy across numerous Natural Language Processing (NLP) tasks, including sentiment analysis. These models benefit from extensive pretraining on diverse textual corpora, allowing them to adeptly extract sentiments from financial data, as depicted in \textbf{Figure \ref{fig:overview}}.

Despite their potential, LLMs face challenges when applied to financial sentiment analysis. One significant hurdle is the disconnect between the models' pretraining objectives and the specific requirements of this field. This issue can be mitigated through targeted instruction tuning, which refines model capabilities to enhance financial sentiment prediction \cite{wei2021finetuned}. Additionally, the inherent brevity of sources such as news flashes and tweets often restricts the available contextual information. To address this, Retrieval-Augmented Generation (RAG) techniques have been introduced, enriching LLMs with external information during inference to improve accuracy and reliability \cite{lewis2020retrieval}.

Furthermore, recent research by Zhang et al. \cite{zhang2023enhancing} introduced a retrieval-augmented framework for financial sentiment analysis that utilizes multi-source queries and keyword similarity-based retrieval. This approach enhances model predictions by leveraging diverse and qualitatively varied information sources. However, reliance on token similarity metrics can introduce biases, potentially favoring contexts that are misleading and fabricated. This underscores the need for careful weighting of different knowledge sources within the RAG module.

The primary objective of financial sentiment analysis is the development of trading strategies and the guidance of portfolio allocation, as illustrated in \textbf{Figure \ref{fig:overview}}. Therefore, integrating market feedback—specifically, aligning sentiment predictions with subsequent market returns—demonstrates a potential synergy between extracting underlying sentiment and predicting stock movements. Reflecting on this market feedback, we refine the weighting of knowledge sources in our financial LLMs' RAG module through direct adjustment or a reinforcement learning (RL)-based approach. This weight adaptation aims to optimize the contributions from various knowledge sources within the RAG module, thereby improving the overall efficacy of financial sentiment analysis. Meanwhile, using more accurate sentiment outputs as trading signals can, in turn, help predict stock movements.

Moreover, prevailing research has largely focused on LLMs of a smaller scale, approximately 7 billion parameters. The concept of emergent abilities \cite{wei2022emergent} describes how increasing model sizes unveil new capabilities in areas such as arithmetic, language comprehension, and semantic analysis. Larger models, benefiting from extensive training on broader datasets, are anticipated to offer superior contextual understanding. This potential for improved performance in tasks such as sentence classification, in our case, sentiment analysis, invites further exploration into the effects of scaling model size. This paper addresses key research questions (RQs) vital for advancing financial sentiment analysis:

\begin{itemize}
\item \textbf{RQ 1}: How can we enhance the LLMs' RAG module with adaptive and non-uniform weighting of multiple knowledge sources and update the weights based on real-world market feedback?
\item \textbf{RQ 2}: What is the impact of increasing model size on the performance of LLMs in financial sentiment analysis?
\item \textbf{RQ 3}: Does aligning LLMs with market feedback result in sentiment predictions that more accurately reflect near-future stock price movements?
\end{itemize}

To address these RQs, we explore direct weight refinement and RL methods to fine-tune knowledge source inputs in our LLMs framework, guided by market feedback. Our extensive evaluation, encompassing a range of benchmarks and comparisons with various model sizes and cutting-edge LLMs, reveals that our advanced financial LLMs markedly outshine existing solutions in financial sentiment analysis. Our primary contributions include:

\begin{itemize}
\item The development of an innovative retrieval-augmented LLM framework that dynamically modulates weights in response to market feedback. This framework introduces a suite of financial LLMs finetuned on foundational models like LLaMA 2 \cite{touvron2023llama2}, extending from pretrained to complex models enriched with Instruction Tuning and RAG, further optimized by both direct feedback and RL techniques. A detailed exposition of these models is provided in \textbf{Section \ref{sec:methodology}}.
\item The superiority of our leading model over the most recent conversational AI systems and the results of recent research in terms of accuracy and F1 score across evaluated benchmarks. We also observe a steady, though modest, enhancement in performance as LLM sizes increase from 7B to 70B parameters. Furthermore, we display the distribution of weights among the different knowledge sources consulted by the LLMs, providing a clearer understanding of the information retrieval mechanism.
\item This analysis confirms the effectiveness of our LLMs in forecasting stock price movements. By implementing long-short strategies derived from sentiment outputs in both bullish and bearish markets, our best model showcases a superior Sharpe ratio when compared to the S\&P 500 portfolio.
\end{itemize}

The paper is organized as follows: \textbf{Section \ref{sec:relatedwork}} reviews the background and related work. \textbf{Section \ref{sec:methodology}} describes how to finetune financial LLMs and methods for refining knowledge source weights. \textbf{Section \ref{sec:results}} details our evaluation metrics, experimental setups, and findings. \textbf{Section \ref{sec:conclusions}} concludes the work and suggests directions for future research.

\section{Related Work}
\label{sec:relatedwork}

\subsection{Methods in Financial Sentiment Analysis}

Sentiment analysis is a critical aspect of financial NLP, aiding in the interpretation of market sentiments to inform investment strategies. Initial studies employed deep learning techniques, LSTM, to handle sequential data and understand financial narratives effectively \cite{sohangir2018big}. The advent of BERT \cite{devlin2018bert} revolutionized NLP by providing a context-aware pretrained model. Adaptations of BERT for finance, such as FinBERT \cite{liu2021finbert, araci2019finbert, yang2020finbert}, have excelled in analyzing financial sentiments by navigating the intricacies of financial language. However, these methods face challenges like insensitivity to numbers, difficulty in discerning sentiment without clear context, and dealing with financial jargon and multilingual content in the face of limited labeled data. The RoBERTa model \cite{liu2019roberta} is an extension and modification of the original BERT model. Due to its extended training on larger datasets and dynamic masking, RoBERTa is better at understanding context and nuances in language.

Recently, LLMs have emerged as a potent solution, offering advanced in-context learning and reasoning capabilities. They facilitate zero-shot learning, performing tasks without domain-specific training \cite{brown2020language, openai2023gpt4}. BloombergGPT \cite{wu2023bloomberggpt} represents a significant stride in this direction, tailor-made for financial applications. Yet, its proprietary nature and the high resource demand for training highlight the accessibility and scalability issues of such specialized models. Consequently, recent research has shifted towards fine-tuning open-source foundation models like LLaMA \cite{touvron2023llama1, touvron2023llama2} to broaden access to efficient financial LLMs.

Despite these advancements, LLMs tailored for finance still face challenges in accurate sentiment prediction due to the mismatch between their general training goals and the specific needs of financial sentiment analysis. This problem is particularly pronounced when analyzing brief content like news flashes and tweets, where limited background information impedes the LLMs' ability to correctly assess sentiments.

\subsection{Instruction Tuning}

Aligning LLMs with specific user instructions presents a notable challenge. Instruction tuning has been identified as an effective strategy for addressing this, enhancing LLMs to better respond to human feedback \cite{wei2021finetuned}. This approach involves training models with \textbf{\textit{(INSTRUCTION, OUTPUT)}} pairs—where \textbf{\textit{INSTRUCTION}} defines the task for the model and \textbf{\textit{OUTPUT}} is the model's expected response based on the instruction \cite{ouyang2022training}. Such tuning improves LLMs' performance in financial tasks by enabling them to understand and execute complex instructions more accurately. Recent research \cite{yang2023investlm} adapts instruction tuning for the financial domain, demonstrating the effectiveness of tailoring LLMs to comprehend and perform tasks in investment analysis and decision-making. To make instruction tuning more efficient, Parameter Efficient Fine-Tuning (PEFT) techniques such as Low-rank Adaptation (LoRA) \cite{hu2021lora} and Quantized Low-rank Adaptation (QLoRA) \cite{dettmers2023qlora} have been developed. These methods optimize fine-tuning by significantly reducing the number of trainable parameters, thereby making the training process more cost-effective. This innovation allows for the more practical application of LLMs in specialized fields like financial sentiment analysis without the need for extensive computing resources.

\subsection{Retrieval Augmented Generation}

Retrieval-augmented generation (RAG) merges the strengths of LLMs with the capabilities of dynamic, external databases to enhance model outputs with up-to-date, domain-specific information \cite{lewis2020retrieval}. This method utilizes external knowledge sources as a form of non-parametric memory, seamlessly integrating retrieved data with the model's input prompt to produce enriched outputs. RAG has proven effective in complex NLP tasks, such as open-domain question answering, showcasing its ability to augment LLMs' responses with relevant context \cite{lewis2020retrieval}. Recent advancements have tailored RAG for financial sentiment analysis, leveraging it to tackle the challenges posed by the concise and rapidly evolving nature of financial news \cite{zhang2023enhancing}. By incorporating relevant and current information from external sources, LLMs enhanced with RAG are able to provide more accurate sentiment analysis, reducing the occurrence of hallucinations \cite{ji2023survey}. This approach is crucial for developing reliable and robust applications of LLMs in the financial domain, ensuring that sentiment analysis is reflective of the most current market conditions and insights.

\subsection{Enhancing LLMs through Feedback}

The role of feedback in the iterative improvement of LLMs has garnered increasing attention in recent research, emphasizing the models' ability to evolve and adapt through feedback integration. One approach is for LLMs to self-improve by internally generating feedback loops. The main concept involves generating an initial output with an LLM; then, the same LLM evaluates its output and uses this feedback to refine itself iteratively \cite{madaan2024self}. Beyond internal mechanisms, incorporating real-world feedback into LLMs offers a valuable approach for real-time adaptation and learning. In the case of QuantAgent \cite{wang2024quantagent}, responses are tested in real-world scenarios to enrich the knowledge source with new insights, requiring minimal human intervention. Recent research by Lopez et al. reveals a significant relationship between market sentiments analyzed by ChatGPT and subsequent stock market behaviors, suggesting that aligning sentiments with real market returns can serve as effective feedback for guiding LLMs \cite{lopez2023can}.

Furthermore, Reinforcement Learning (RL) has been recognized as a potent tool for refining the decision-making capabilities of LLMs. Inspired by Reinforcement Learning from Human Feedback (RLHF) \cite{ouyang2022training}, researchers have investigated various RL algorithms aimed at enhancing LLMs by learning from external feedback, thereby bolstering their reasoning processes \cite{havrilla2024teaching}. AdaRefiner introduces a novel framework to enhance the synergy between LLMs and RL feedback, contributing to the automatic self-refinement of LLMs with RL feedback. By leveraging feedback loops, real-world interactions, and reinforcement learning strategies, LLMs can achieve greater accuracy, adaptability, and relevance across a range of applications.

\section{Methodology}
\label{sec:methodology}

\subsection{Overview}
\label{sec:overview}

\begin{figure*}[ht!]
\centering
\includegraphics[width=0.95\textwidth]{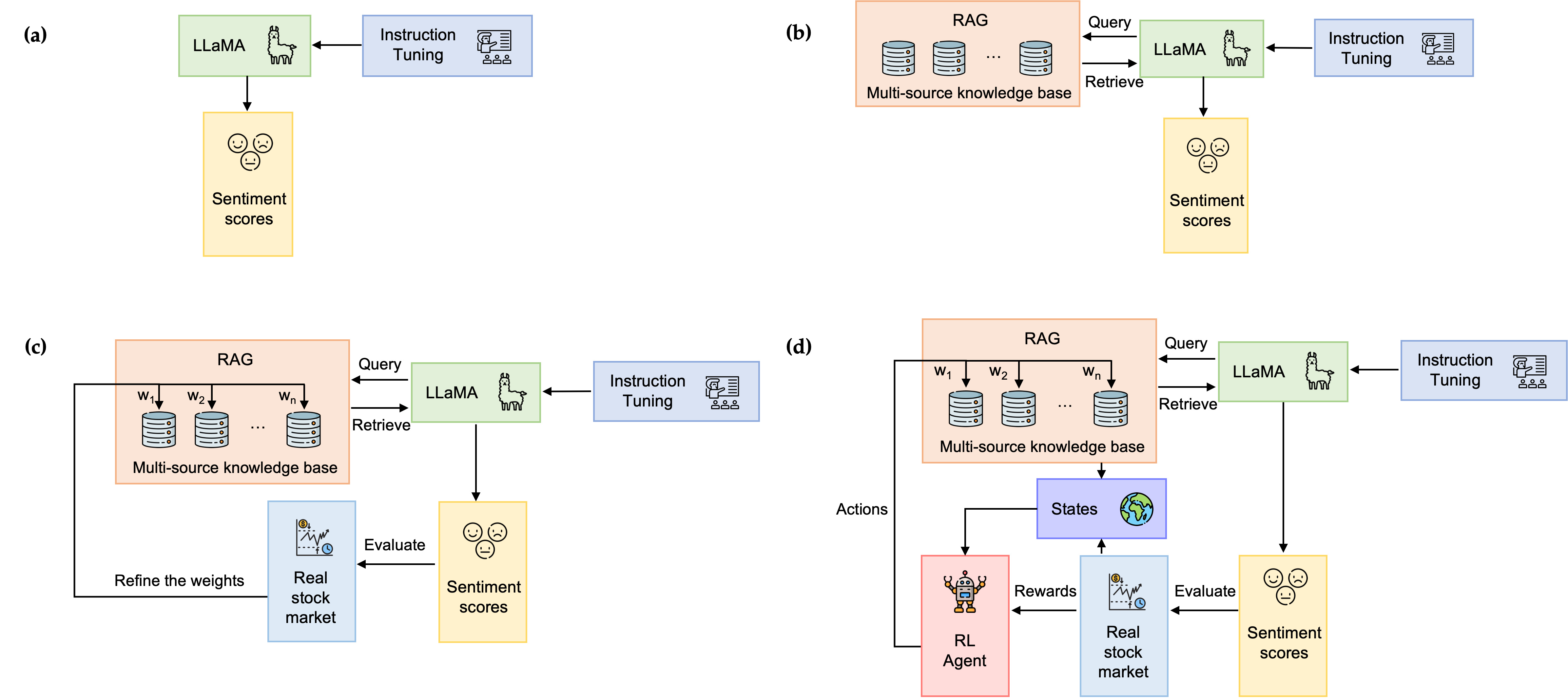}
\caption{Family of financial LLMs based on the LLaMA 2. \textnormal{(a) LLaMA I, (b) LLaMA I-RAG, (c) LLaMA I-RAG-DF, (d) LLaMA I-RAG-RL.}}
\label{fig:methods}
\end{figure*}

We present a novel, adaptive retrieval-augmented LLMs framework that integrates market feedback for more dynamic weight allocation across different knowledge sources within the RAG module. Building on the LLaMA 2 architecture \cite{touvron2023llama2}, we introduce a suite of financial LLMs tailored for financial sentiment analysis, as depicted in \textbf{Figure \ref{fig:methods}}, which includes:

\begin{itemize}
\item \textbf{LLaMA P}: Pretrained LLaMA 2 models without any modifications.
\item \textbf{LLaMA I}: Enhanced LLaMA 2 models incorporating instruction tuning for an improved understanding of human instructions.
\item \textbf{LLaMA I-RAG}: LLaMA 2 models equipped with instruction tuning and a standard RAG for retrieving external information.
\item \textbf{LLaMA I-RAG-DR}: LLaMA 2 models that further refine the RAG module's weights directly based on market feedback, in addition to instruction tuning.
\item \textbf{LLaMA I-RAG-RL}: The most advanced iteration, combining LLaMA 2 models with instruction tuning and an adaptive RAG module, where weights are optimized through reinforcement learning techniques.
\end{itemize}

The following sections will delve into the details of each component and methodological approach.

\subsection{Instruction Tuning}

To implement instruction tuning for the LLaMA 2 series, we adopted a methodology analogous to the Instruct-FinGPT approach \cite{zhang2023instruct}. Our initial step involved constructing an instruction dataset from existing financial sentiment analysis datasets at minimal cost. We crafted ten distinct yet related instructions to elucidate the task of financial sentiment analysis, with details provided in \textbf{Appendix Section \ref{sec:instruction}}. For each data sample, we paired a randomly selected instruction with its corresponding input and output. This pairing was formatted as: "Human: \textbf{\textit{INSTRUCTION} + \textit{INPUT}}, Assistant: \textbf{\textit{OUTPUT}}." Subsequently, we fine-tuned the LLMs using the causal language modeling (CLM) objective, which enhances the prediction of the next token based on prior context. Despite these efforts, the model occasionally generated extraneous free-style text that did not correspond to the targeted sentiment labels, a phenomenon observed both in the pretrained and some fine-tuned instances. In such cases, we assigned the predominant keyword to one of three predefined sentiments: positive, negative, or neutral. If this approach proved ineffective, manual verification was employed, although this was necessary in only approximately 1\% of cases.

\subsection{RAG Module}

We construct an external multi-source knowledge source incorporating various sources, as detailed in \textbf{Table \ref{table:financial-sources}}. The selection criteria for these sources include the availability of automated retrieval APIs, reliability and authority, relevance to finance, data quality and format, and diversity of perspectives. These sources provide additional context for LLMs to generate sentiments. Specifically, we apply non-uniform weights $\left(w_1, w_2, \ldots, w_K\right)$ to each source and ensure these weights are normalized to sum to 1.

\begin{table*}[ht!]
\centering
\begin{tabular}{l|l|l|l}
\hline
\textbf{Source} & \textbf{Type} & \textbf{Primary Focus} & \textbf{Accessibility} \\
\hline
Bloomberg & News & Financial markets, economic news & Subscription-based \\
CNBC & News & Financial markets, business news & Free, with premium options \\
Morningstar & Research Publication & Investment research, stock ratings & Subscription-based \\
Reddit & Social Media & Public discussions, market sentiments & Free \\
Seeking Alpha & Research Publication & Company analysis, investment strategies & Free, with premium options \\
The Wall Street Journal & News & Business news, financial markets & Subscription-based \\
Twitter & Social Media& Real-time news, public sentiment & Free \\
Yahoo Finance & News & Stock market news, financial reports & Free, with premium options \\
\hline
\end{tabular}
\caption{Selected knowledge sources for RAG module.}
\label{table:financial-sources}
\end{table*}

We process original statements intended for sentiment analysis as queries by using regular expressions to extract a set of relevant keywords as queries. Then we use platform internal retrieval APIs to gather a batch of documents daily, forming a candidate pool that is pre-filtered by stock symbol and date. Then, the Weighted Overlap Coefficient (WOC) is employed to consider the weight of each source $w_i$ and the overlap between the query and the document.

\begin{equation}
    W O C(X, Y, w)=w \times \frac{|X \cap Y|}{\min (|X|,|Y|)}
\label{eq:woc}
\end{equation}

 \noindent where $X$ and $Y$ represent sets of relevant tokens \footnote{We use the NLTK library for Named-entity recognition (NER) and remove words with neutral sentiment before performing tokenization.} from the queried sentence and its context, respectively, and $w$ is the weight corresponding to the source of $Y$. The top $K$ documents, determined by WOC, are appended to the original query, along with instructions, to derive the final sentiment output. The detailed algorithm is described in \textbf{Algorithm \ref{algo:RAG_WOC}}. A detailed, illustrated example of how the RAG module works is shown in \textbf{Appendix Section \ref{sec:details}}.

\newcommand{\INPUT}[1]{\STATE \textbf{Input:} #1}
\newcommand{\OUTPUT}[1]{\STATE \textbf{Output:} #1}

\begin{algorithm}
\caption{RAG with Weighted Overlap Coefficient (WOC)}
\begin{algorithmic}[1]
\INPUT{Query $Q$, Weights of knowledge source $W$}
\OUTPUT{Context $C$}
\STATE Initialize $C \leftarrow \emptyset$
\STATE Retrieve a candidate pool of documents $D$, each potentially relevant to $Q$
\STATE Initialize an empty list $P$ to store documents and WOC scores
\FOR{each document $d \in D$}
\STATE Obtain the weight $w_j$ corresponding to the source of $d$
\STATE Calculate $WOC(Q, d, w_j)$
\STATE Add $(d, WOC(Q, d, w_j))$ to $P$
\ENDFOR
\STATE Sort $P$ in descending order based on WOC scores
\STATE Select the top $K$ entries from $P$ and add their documents to $C$
\STATE \textbf{return} $C$
\end{algorithmic}
\label{algo:RAG_WOC}
\end{algorithm}

\subsection{Direct Weights Refinement by Market Feedback}
\label{sec:DR}

In line with our discussion in \textbf{Section \ref{sec:relatedwork}}, we enhance our model by comparing our sentiment predictions with actual stock returns. We calculate the daily return of a stock on day $T$ as the percentage change in its open price from day $T-1$ to day $T$, indicating its stock price movement. Stock movements within $\pm 1$ standard deviation from the mean, considered as the range for neutral movements, account for about $68\%$ of daily price changes under a normal distribution assumption. Every day, we look at both the sentiment score and the actual stock return. We deem a prediction for day $T$ as accurate if it meets any of the following criteria:
\begin{itemize}
    \item The return on day $T+1$ is positive and falls outside the neutral range, matching a sentiment output "positive";
    \item The return on day $T+1$ is negative and falls outside the neutral range, matching a sentiment output "negative";
    \item The return on day $T+1$ is within the neutral range, matching a sentiment output "neutral".
\end{itemize}

If a prediction doesn't fit these scenarios, we mark it as inaccurate, which prompts a negative feedback loop. We then adjust the weights of the knowledge sources used as follows:

For accurate predictions:

\begin{equation}
w_i^{\text{new}} = w_i^{\text{old}} + \alpha
\end{equation}

And for inaccurate ones:

\begin{equation}
w_i^{\text{new}} = w_i^{\text{old}} - \alpha
\end{equation}

In the formulas above, $i \in K$, we only adjust the weights for sources we actually feed into LLMs; the rest remain unchanged. The constant parameter $\alpha$, which influences how much we adjust the weights, is set to 1e-4 in our experiments. We aggregate the weight changes and update weights on a batch basis. The specifics of this approach are detailed in \textbf{Algorithm \ref{algo:DF}}.

\begin{algorithm}
\caption{Direct Refinement of Knowledge Source Weights by Market Feedback}
\begin{algorithmic}[1]
\INPUT{Initial uniform weights $W = \{w_1, w_2, \ldots, w_k\}$, Source usage per feedback $S = \{s_1, s_2, \ldots, s_n\}$, Market feedback $F = \{f_1, f_2, \ldots, f_n\}$, Learning rate $\alpha > 0$, Batch size $B$}
\OUTPUT{Refined weights $W'$}
\FOR{$b = 1$ to $B$}
    \STATE Initialize $\Delta W = \{0, 0, \ldots, 0\}$ with length $k$ to store weight changes for each batch
    \FOR{$i = 1$ to $n$ within batch $b$}
        \FOR{each source $j \in s_i$}
            \IF{$f_i$ indicates a correct prediction}
                \STATE $\Delta w_j \leftarrow \Delta w_j + \alpha$
            \ELSE
                \STATE $\Delta w_j \leftarrow \Delta w_j - \alpha$
            \ENDIF
        \ENDFOR
    \ENDFOR
    \FOR{$j = 1$ to $k$}
        \STATE $w_j \leftarrow \max(0, w_j + \Delta w_j)$ to ensure non-negative
    \ENDFOR
    \STATE Normalize $W'$ such that $\sum_{j=1}^{k} w_j' = 1$
\ENDFOR
\RETURN $W'$
\end{algorithmic}
\label{algo:DF}
\end{algorithm}

\subsection{RL-based Weights Refinement by Market Feedback}
\label{sec:RL}

In this section, we describe how we refine the weights of knowledge sources in the RAG module using Reinforcement Learning (RL), opting for a more dynamic approach compared to direct adjustments. This method allows for a more sophisticated and potentially more effective mechanism for adapting weights based on feedback, leveraging the strengths of RL in dealing with continuous action spaces and optimizing long-term rewards.

Our RL setup features a continuous action space where actions, represented as a vector $\mathbf{w}=$ $\left(w_1, w_2, \ldots, w_K\right)$, detail the weights for the knowledge sources. We constrain each weight $w_i$ to fall within $[0,1]$, ensuring the total weight set sums to 1. This approach builds on the similar market feedback mechanism outlined in \textbf{Section \ref{sec:DR}}, with a reward function that assigns +1 for correct sentiment predictions and -1 for incorrect ones. The state for our RL model captures the current condition of RAG along with relevant market information, such as current weights, historical accuracy rate, average overlap coefficient with the query across different knowledge sources, and historical stock return and technical indicators (more details are shown in \textbf{Appendix Section \ref{sec:state}}).

In our study, we adopt Proximal Policy Optimization (PPO) \cite{schulman2017proximal}, a reinforcement learning algorithm noted for its stability, efficiency, and ease of implementation, qualities particularly valuable in the volatile environment of stock trading. PPO is a type of policy gradient method, which optimizes decision-making policies directly and is well-suited for scenarios with large or continuous action spaces.

PPO addresses a common challenge in training policy gradient methods: significant policy degradation due to large, destabilizing updates. It achieves this by introducing a novel objective function that limits the size of policy updates, promoting incremental learning and preventing large shifts that could negatively affect performance. This moderation is particularly crucial in maintaining the delicate balance between exploration of new strategies and exploitation of known profitable behaviors.

The core mechanism of PPO involves a probability ratio, \( r_t(\theta) = \frac{\pi_\theta(a_t | s_t)}{\pi_{\theta_{old}}(a_t | s_t)} \), which compares the likelihood of taking an action under the new policy versus the old policy. This ratio is incorporated into PPO's clipped objective function:

\[
J^{CLIP}(\theta) = \hat{\mathbb{E}}_t\left[\min \left(r_t(\theta) \hat{A}(s_t, a_t), \operatorname{clip}(r_t(\theta), 1-\epsilon, 1+\epsilon) \hat{A}(s_t, a_t)\right)\right]
\]

In this equation, \( \hat{A}(s_t, a_t) \) represents the advantage function, estimating the relative benefit of choosing a particular action compared to the baseline. The clipping operation restricts the ratio \( r_t(\theta) \) within the range \([1-\epsilon, 1+\epsilon]\), controlling the extent of policy updates and thus enhancing the stability of the training process. For more details, please refer to the original PPO paper \cite{schulman2017proximal}.

Our PPO-based reinforcement learning architecture consists of several fully connected layers designed to process state inputs effectively. The architecture is divided into two primary components: a policy network and a value network. The policy network is responsible for determining the action outputs, specifically the allocation of weights across \( K \) knowledge sources within the RAG module. Instead of using a softmax activation layer typically suited for discrete action spaces, we employ a linear activation function in the final dense layer of the policy network to produce a continuous action vector. A post-processing normalization step ensures that these weights sum to one. Concurrently, the value network, branching from a shared dense layer, features a single output neuron with linear activation to estimate the state's value, facilitating an assessment of the current policy's expected long-term returns. Details of this method are fully laid out in \textbf{Algorithm \ref{algo:RAG_RL}}.

\begin{algorithm}
\caption{RL-based Refinement of RAG Weights by Market Feedback using PPO}
\begin{algorithmic}[1]
\INPUT{Initial normalized weights $W = \{w_1, w_2, \ldots, w_K\}$, State features: current weights, historical accuracy, overlap coefficients, market indicators}
\OUTPUT{Refined weights $W'$}
\STATE Initialize policy network $\pi_\theta$ for predicting weight adjustments
\STATE Initialize value network $V_\phi$ for state value estimation
\WHILE{not converged}
    \FOR{each iteration}
        \STATE Observe state $s$ (including market indicators and performance metrics)
        \FOR{each environment interaction within the iteration}
            \STATE Predict action $a = (w_1, \ldots, w_K)$ using $\pi_\theta(s)$
            \STATE Renormalize $a$ to ensure $\sum w_i = 1$ for action execution
            \STATE Apply $a$, observe new state $s'$, and calculate reward $r$ based on prediction against market movements
            \STATE Update state $s \leftarrow s'$ 
        \ENDFOR
        \STATE Compute advantage estimates $\hat{A}$ based on rewards and $V_\phi$ predictions
        \STATE Update $\pi_\theta$ by maximizing PPO objective w.r.t $\theta$
        \STATE Update $V_\phi$ by minimizing value function loss w.r.t $\phi$
    \ENDFOR
\ENDWHILE
\STATE \textbf{Output:} Optimized weights $W'$ from refined policy $\pi_\theta$
\end{algorithmic}
\label{algo:RAG_RL}
\end{algorithm}

\section{Performance Evaluation}
\label{sec:results}

In this section, we assess the efficacy of our proposed suite of financial LLMs for the financial sentiment analysis task. We compare the results with state-of-the-art approaches from previous studies and current cutting-edge conversational AI systems. To assess the effectiveness of each model, we report accuracy and F1 scores on benchmark datasets.

\subsection{Datasets}

To fairly compare results with existing research \cite{zhang2023instruct, zhang2023enhancing, fatemi2023comparative}, we used the same public datasets that are accessible through HuggingFace for training and testing the model. Specifically, we use the following datasets to train and run inference for the LLaMA I and LLaMA I-RAG models, boosting performance by instruction tuning and RAG technique.

The training dataset is a combination of the training split of Twitter Financial News Sentiment (TFNS)\footnote{\url{https://huggingface.co/datasets/zeroshot/twitter-financial-news-sentiment}} and the FiQA SA dataset\footnote{\url{https://huggingface.co/datasets/pauri32/fiqa-2018}}, resulting in a total of 10,504 samples. This combined dataset was adapted to fit an instruction-following format suitable for instruction tuning. For model evaluation, we utilized the validation segment of the TFNS dataset alongside the Financial PhraseBank (FPB) dataset\footnote{\url{https://huggingface.co/datasets/financial_phrasebank}}, totaling 7,234 samples. A notable challenge with these datasets is their frequent omission of stock symbols and specific date tags—information crucial for maximizing the RAG module's retrieval capabilities.

For the LLaMA I-RAG-DF and LLaMA I-RAG-RL models, the term "train" refers to the process of applying techniques outlined in \textbf{Section \ref{sec:DR}} and \textbf{\ref{sec:RL}} to fine-tune RAG weights, rather than directly modifying the LLMs' model weights. This process necessitates the inclusion of real market feedback, demanding specific data such as timestamps and stock symbols. To meet this requirement, we manually created a specialized dataset comprising 20,000 news headlines, with each randomly selected stock and date ranging from 01/01/2018 to 12/31/2020. Each headline is annotated with relevant temporal and stock symbol information, alongside the subsequent day's stock return. Performance metrics for these models were then evaluated using the same validation sets from TFNS Val and FPB as the other models for consistency.

\begin{table*}[htbp!]
\centering
\begin{tabular}{lcccc}
\hline
\multicolumn{1}{c}{Model} & \multicolumn{2}{c}{FPB} & \multicolumn{2}{c}{TFNS (Val)} \\
 & Accuracy & F1 & Accuracy & F1 \\
\hline
\multicolumn{5}{l}{\textbf{Our financial LLMs (70B)}} \\
\hline
LLaMA P & 0.652 & 0.425 & 0.605 & 0.408 \\
LLaMA I & 0.738 $\pm$ 0.018 & 0.718 $\pm$ 0.018 & 0.834 $\pm$ 0.016 & 0.765 $\pm$ 0.015 \\
LLaMA I-RAG & 0.781 $\pm$ 0.019 & 0.741 $\pm$ 0.012 & 0.883 $\pm$ 0.016 & 0.819 $\pm$ 0.010 \\
LLaMA I-RAG-DF & 0.812 $\pm$ 0.018 & 0.782 $\pm$ 0.022 & 0.905 $\pm$ 0.017 & 0.870 $\pm$ 0.015 \\
LLaMA I-RAG-RL & 0.810 $\pm$ 0.024 & 0.781 $\pm$ 0.021 & 0.910 $\pm$ 0.021 & 0.878 $\pm$ 0.019 \\
\hline
\multicolumn{5}{l}{\textbf{Pre-trained language model}} \\
\hline
RoBERTa & 0.723 $\pm$ 0.021 & 0.689 $\pm$ 0.011 & 0.804 $\pm$ 0.014 & 0.741 $\pm$ 0.009 \\
\hline
\multicolumn{5}{l}{\textbf{Closed-sourced LLMs}} \\
\hline
BloombergGPT & - & 0.511 & - & - \\
GPT-3.5 Turbo & 0.654 & 0.529 & 0.673 & 0.484 \\
GPT-4 Turbo & 0.773 & 0.702 & 0.835 & 0.794 \\
Gemini Pro & 0.725 & 0.673 & 0.809 & 0.775 \\
Claude-3 & 0.759 & 0.697 & 0.812 & 0.786 \\
\hline
\multicolumn{5}{l}{\textbf{Recent LLMs research}} \\
\hline
Zhang et al. (2023) \cite{zhang2023enhancing} & 0.758 & 0.739 & 0.863 & 0.811 \\
Fatemi \& Hu (2023) \cite{fatemi2023comparative} & 0.807 & 0.780 & 0.903 & 0.878 \\
\hline
\end{tabular}
\caption{Comparative Performance of different LLMs. \textnormal{The error terms represent the standard deviation calculated across ten independent training experiments, each with a different random seed.}}
\label{tab:llms_performance}
\end{table*}

\subsection{Model Training}

For efficient parameter fine-tuning of LLMs, we employed Low-Rank Adaptation (LoRA) \cite{hu2021lora}, setting the rank to 8 and the alpha value to 32. We specifically targeted LoRA modules such as "q\_proj", "k\_proj", "v\_proj", "down\_proj", "gate\_proj", and "up\_proj". Our training utilizes the AdamW optimizer, with a training epochs of 10, a batch size of 32, a starting learning rate of 5e-5, and a weight decay rate of 0.1. To ensure efficiency, particularly given the LLaMA 2 model's maximum context capacity of 4,096 tokens, we implemented a practical token limit of 2,048 for each input sample. All models undergo training using bfloat16 mixed-precision on a robust hardware setup of eight A100 (40GB) GPUs at Lambda Labs\footnote{\url{https://lambdalabs.com/service/gpu-cloud}}, completing within a few hours. 

For fine-tuning the RoBERTa model used in \textbf{Section \ref{sec:model_res}}, we downloaded the pre-trained RoBERTa Large model from Hugging Face\footnote{\url{https://huggingface.co/FacebookAI/roberta-large}}. A classification head was appended to tailor it for categorizing text into sentiment classes. This model was fine-tuned on the exact same benchmark datasets using the AdamW optimizer, with a learning rate of 2e-5 and a batch size of 16, running for 4 epochs to ensure precise convergence and prevent overfitting.

In our Proximal Policy Optimization (PPO) configuration, we set a policy learning rate of 5e-4, an update frequency of 10 epochs, a discount factor of 1 for immediate reward evaluation, and a clip ratio of 0.2 to moderate policy updates. It is worth noting that, while we selected PPO for its balance of efficiency and performance, the design of our LLaMA I-RAG-RL model is compatible with a variety of reinforcement learning algorithms, not limited to PPO. In terms of potential multiple sources of randomness such as model initialization, data mini-batch shuffle during training, and random sampling of news to compute sentiment scores, etc., we conducted ten independent experiments under the same setting with respect to different random seeds.

\subsection{Results}

\subsubsection{Model Performance Comparison}
\label{sec:model_res}

To address \textbf{RQ 1} as outlined in \textbf{Section \ref{sec:intro}}, we developed two methods for constructing an adaptive Retrieval-Augmented Generation (RAG) system, as detailed in \textbf{Sections \ref{sec:DR}} and \textbf{\ref{sec:RL}}. To evaluate their empirical performance rigorously in financial sentiment analysis, we benchmarked our LLaMA with 70 billion parameters against widely recognized pre-trained language models, such as RoBERTa Large (335M parameters), and several leading conversational AI systems, including OpenAI's GPT-4-0125-preview \cite{achiam2023gpt}, GPT-3.5-turbo-0125 \cite{ouyang2022training}, Anthropic’s Claude-3-Opus \cite{anthropic2023claude}, and Google’s Gemini-Pro \cite{team2023gemini}. Additionally, we included BloombergGPT \cite{wu2023bloomberggpt}, a first-of-its-kind, finance-focused LLM in our benchmark, relying on its reported metrics on the Financial PhraseBank (FPB) dataset. Our analysis incorporates insights from recent research \cite{zhang2023enhancing, fatemi2023comparative}, with detailed results presented in \textbf{Table \ref{tab:llms_performance}}.

\textbf{Table \ref{tab:llms_performance}} illustrates the progressive performance gains achieved with increasingly sophisticated models. On the TFNS validation dataset, for instance, we observed a 37.85\% improvement in accuracy when LLaMA P was enhanced with instruction tuning to create LLaMA I. The integration of the RAG module further advanced LLaMA I-RAG's accuracy by 5.88\%. The modest uplift from RAG is due to the limited availability of date information and pertinent named entities in tweets and news segments, which constrains the RAG's retrieval capabilities. The LLaMA I-RAG-RL model exhibited an additional 3.06\% gain in accuracy, attributed to reweighting the sources in the RAG module based on their reliability and relevance as indicated by market feedback. Thus, incorporating the feedback that optimizes source weights in the LLMs' RAG module improves sentiment extraction from financial contexts.

On the FPB dataset, there was a notable improvement in accuracy from the transition of LLaMA P to LLaMA I-RAG-DF. It is observed that LLaMA I-RAG-RL did not show a significant improvement over LLaMA I-RAG-DF, suggesting that a simple weight refinement strategy is effective for this dataset.

It is also noteworthy that the RoBERTa model, after fine-tuning, performed only slightly inferior to LLaMA I. Among the closed-source LLMs, GPT-4 Turbo remains the standout performer across the datasets evaluated. However, it is crucial to recognize that classification is merely one facet of NLP capabilities. Before concluding GPT-4 Turbo's superiority over Gemini and Claude, it is imperative to apply benchmarks across more complex tasks such as reasoning, coding, and creative writing to ensure a comprehensive evaluation \cite{chang2023survey}. Our leading model, LLaMA I-RAG-RL, surpasses recent research proposals in both accuracy and F1 score, highlighting the effectiveness of refining the RAG module through direct adjustment or reinforcement learning (RL) methods informed by market feedback.

\subsubsection{Influence of Model Size on Performance}

\begin{figure}[!ht]
\centering
\includegraphics[width=0.48\textwidth]{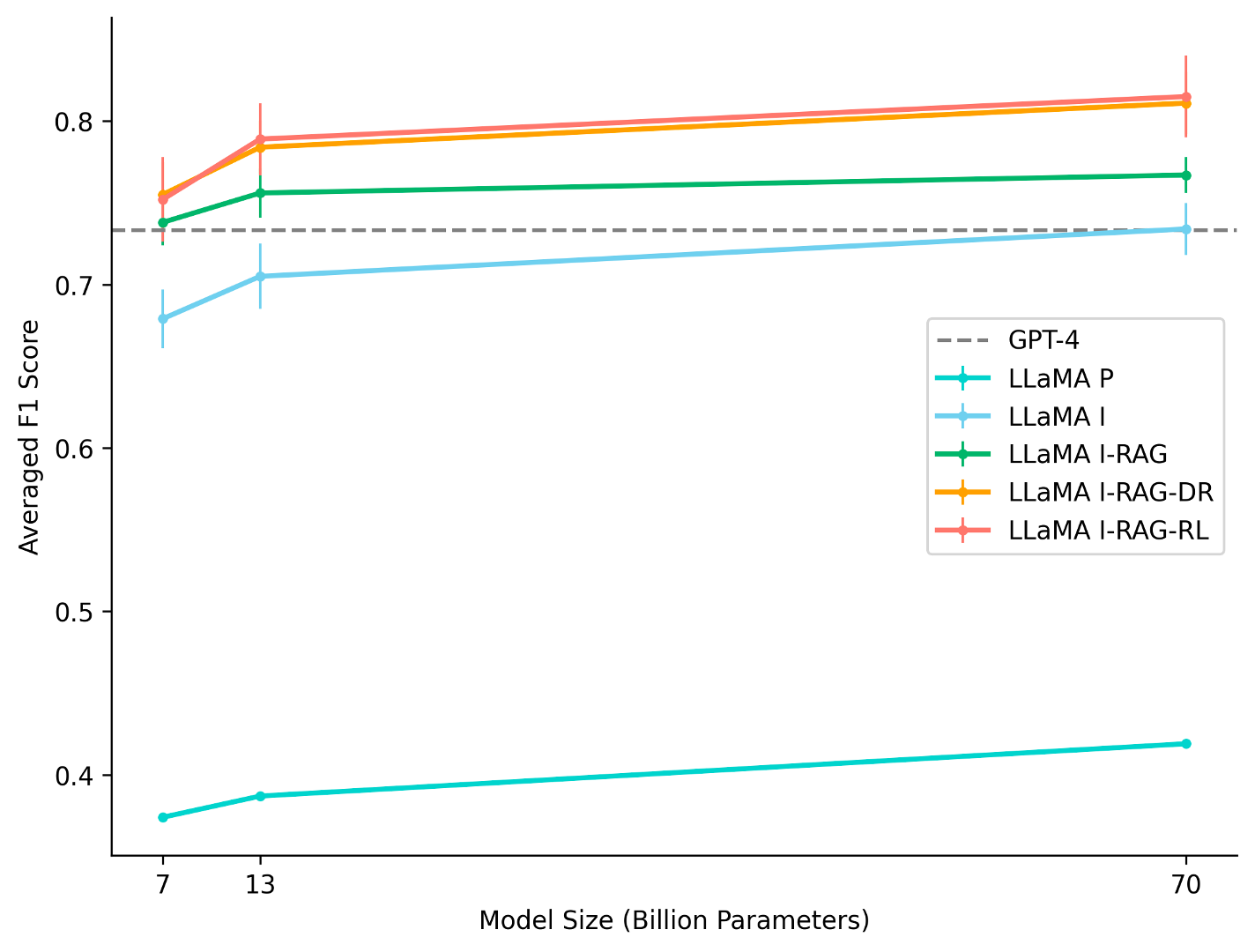}
\caption{The impact of model size on weighted F1 score across test datasets. \textnormal{The length of the error bar extending from this central point throughout the paper represents the standard deviation calculated across ten independent experiments, each with a different random seed. The grey horizontal line indicates the weighted F-1 score obtained by GPT-4 Turbo.}}
\label{fig:model_size}
\end{figure}

Prior studies \cite{zhang2023instruct, zhang2023enhancing, fatemi2023comparative} have primarily evaluated LLMs ranging in size from 250M to 7B parameters. In pursuit of \textbf{RQ 2}, we broadened this scope by training models up to 70B parameters, adhering to the methodologies delineated in \textbf{Section \ref{sec:overview}}. As illustrated in \textbf{Figure \ref{fig:model_size}}, instruction tuning emerged as the most substantial factor in improving F1 scores across various model sizes, consistent with earlier findings. We noted only a slight enhancement in performance when scaling the model size from 7B to 13B, and then to 70B. Interestingly, the 7B variant of the LLaMA I-RAG model attained weighted F1 scores comparable to those of the 70B LLaMA I model, despite having 10 times fewer parameters. Considering the significant increase in computational resources and time required for training larger models, our results suggest that merely expanding model size is less efficient compared to employing techniques like RAG, particularly when dealing with the succinct nature of the financial news segments, headlines, and tweets in our case.

\subsubsection{Influence of Knowledge Source Weights in RAG}

\begin{figure}[!ht]
\centering
\includegraphics[width=0.48\textwidth]{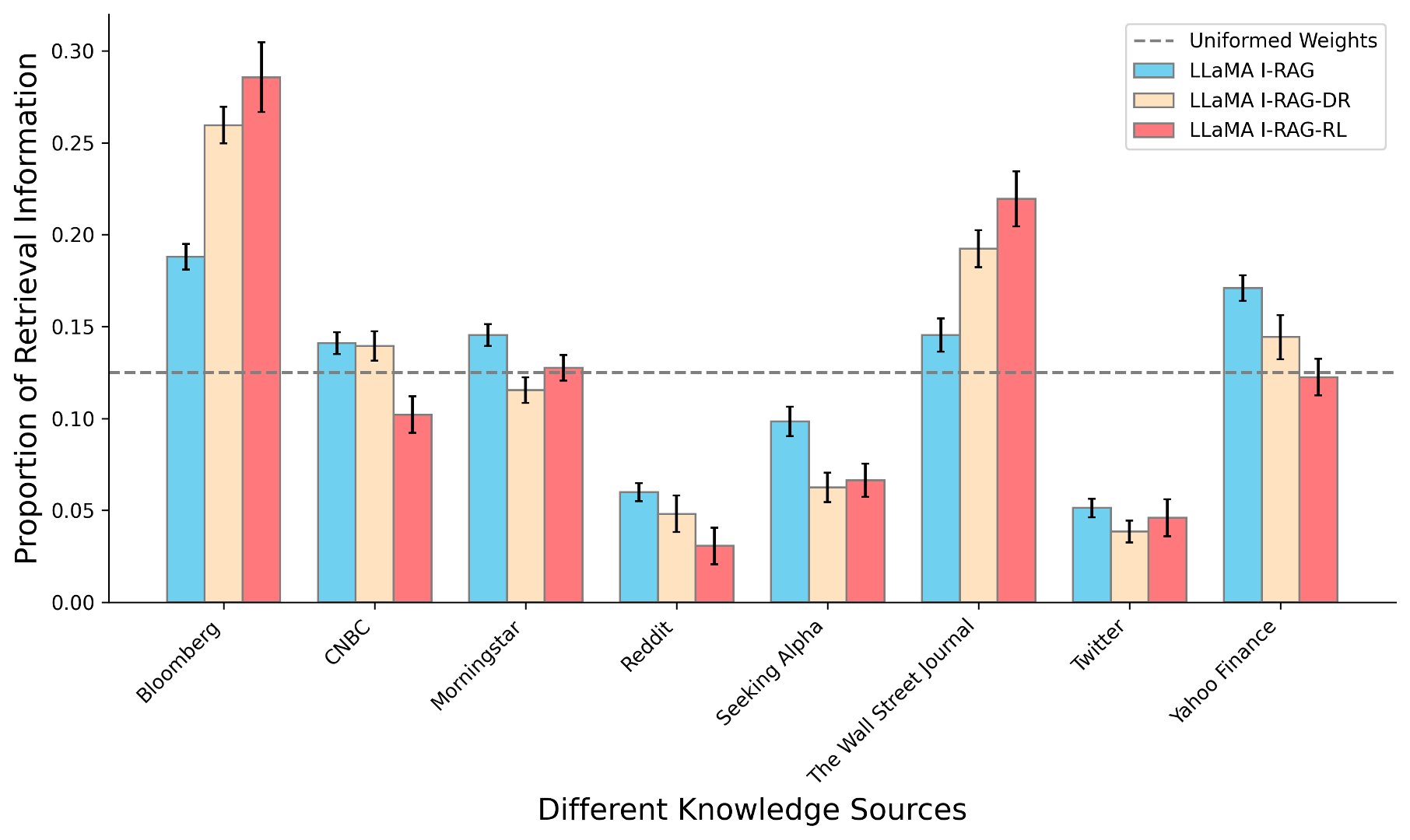}
\caption{Weights distribution across different knowledge sources of RAG. \textnormal{The gray horizontal line represents the uniform initialization level (12.5\%) for the eight evaluated knowledge sources.}}
\label{fig:model_weights}
\end{figure}

This section investigates the impact of applying the direct weights refinement and RL-based weights refinement methods on adapting the source weights within the RAG module. We assessed the final distribution of source weights utilized by our 70B optimized LLMs as follows: For both LLaMA I-RAG-DF and I-RAG-RL models, we employed similarity-based retrieval using \textbf{Equation \ref{eq:woc}} and selected the top K sources with the highest WOC scores. In contrast, for the LLaMA I-RAG model, we uniformly set $w = 1/K$ for all sources. We recorded the retrieval history to track which sources were accessed by each sample and subsequently calculated their proportion by sources. The outcomes are depicted in \textbf{Figure \ref{fig:model_weights}}.

Although the two refinement methods resulted in differing weight distributions, a notable pattern emerged with the RL-based refinement leading to more pronounced imbalances. Specifically, only Bloomberg and The Wall Street Journal surpassed the original uniform weight level (12.5\%) after applying either refinement method, as denoted by the gray horizontal line in the figure. Common to both methods was an increase in the weights assigned to Bloomberg and The Wall Street Journal, ranging between 5\% to 10\%, and a reduction in the weight for socially-driven platforms like Reddit to between 2\% to 3\%. This adjustment aligns with the understanding that postings on Reddit generally possess lower reliability and authority compared to established news platforms. Furthermore, the diverse and broad nature of discussions on Reddit may dilute its correlation with actual stock market movements. Therefore, models adjusted their dependence on such sources based on market feedback. Nevertheless, models still allocated minor weights to sources like Reddit and Twitter, underscoring the importance of source diversity to RAG.

\begin{figure*}[!ht]
    \centering
    \begin{subfigure}[b]{0.68\textwidth}
        \includegraphics[width=\textwidth]{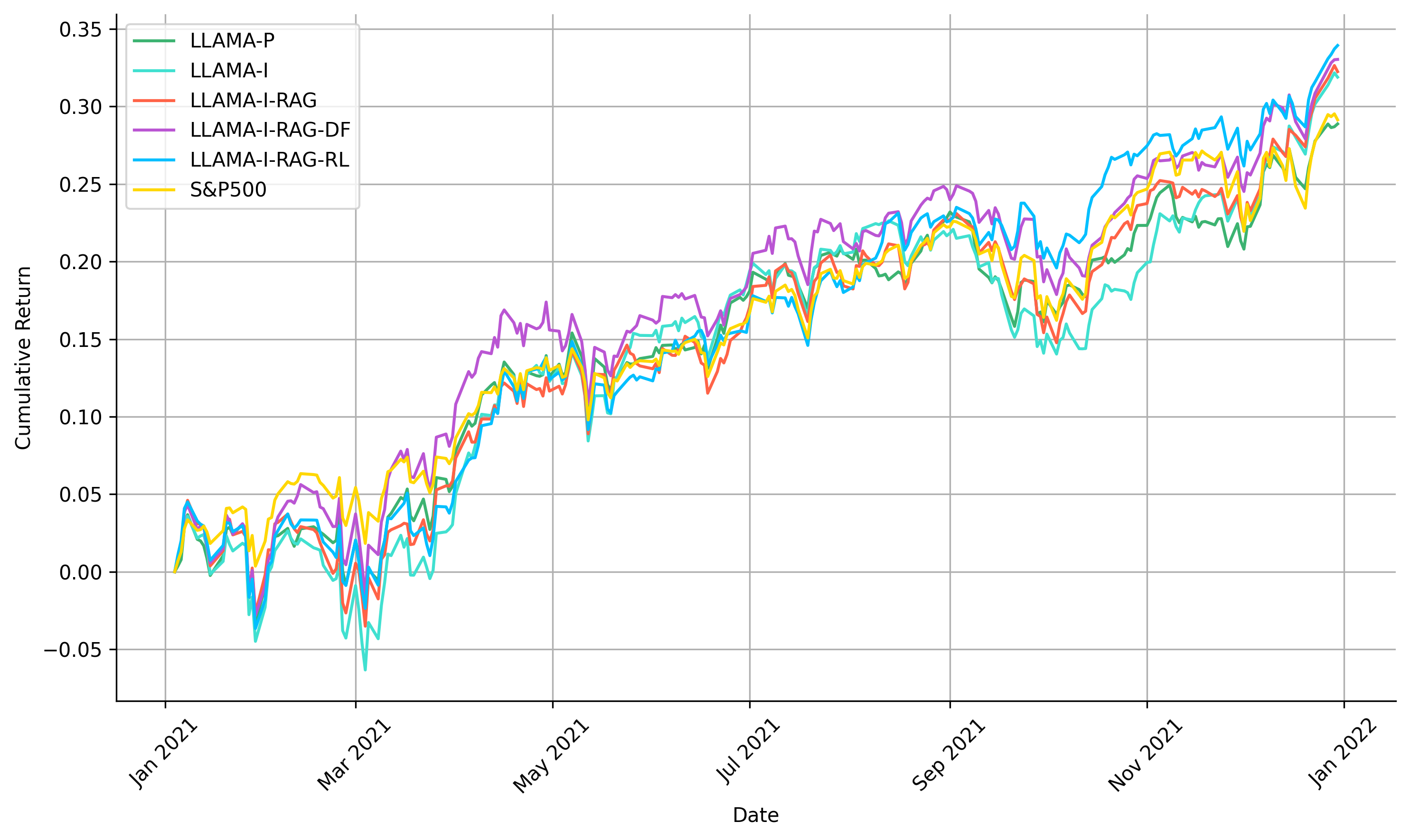}
        \caption{Year 2021 (bullish market)}
        \label{fig:port_2021}
    \end{subfigure}
    
    \vspace{0.5cm} 
    
    \begin{subfigure}[b]{0.68\textwidth}
        \includegraphics[width=\textwidth]{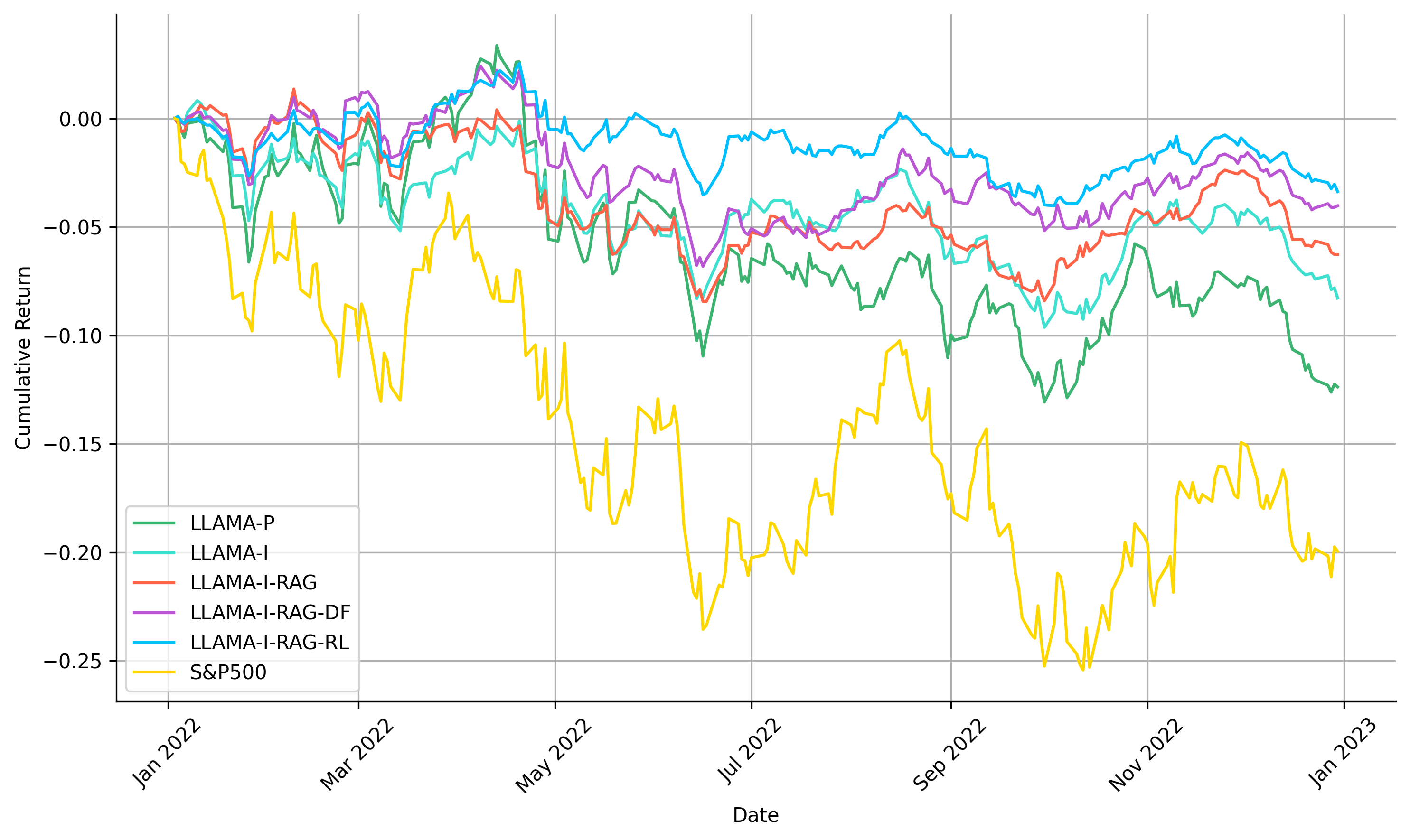}
        \caption{Year 2022 (bearish market)}
        \label{fig:port_2022}
    \end{subfigure}
    
    \caption{Cumulative return curves of different investment strategies and S\&P 500. \textnormal{Values are computed as the mean of ten independent training experiments, each with a different random seed.}}
    \label{fig:model_portfolio}
\end{figure*}

\begin{table*}[ht]
\centering
\caption{Comparative Evaluation of Performance. \textnormal{The error terms represent the standard deviation calculated across ten independent experiments, each with a different random seed.}}
\label{table:performance_comparison}
\begin{subtable}{\textwidth}
\centering
\caption{Year 2021 (bullish market)}
\begin{tabular}{l|cccccc}
\hline
\textbf{Metric} & \textbf{LLaMA P} & \textbf{LLaMA I} & \textbf{LLaMA I-RAG} & \textbf{LLaMA I-RAG-DF} & \textbf{LLaMA I-RAG-RL} & \textbf{S\&P500} \\
\hline
Cumulative Return & 0.2887 $\pm$ 0.008 & 0.3188 $\pm$ 0.009 & 0.3224 $\pm$ 0.007 & 0.3303 $\pm$ 0.010 & 0.3393 $\pm$ 0.011 & 0.2913 \\
Annualized Return & 0.2935 $\pm$ 0.008 & 0.3242 $\pm$ 0.008 & 0.3278 $\pm$ 0.007 & 0.3358 $\pm$ 0.010 & 0.3450 $\pm$ 0.010 & 0.2961 \\
Annualized Volatility & 0.1468 $\pm$ 0.006 & 0.1608 $\pm$ 0.005 & 0.1528 $\pm$ 0.004 & 0.1555 $\pm$ 0.007 & 0.1464 $\pm$ 0.008 & 0.1303 \\
Sharpe Ratio & 1.9990 $\pm$ 0.098 & 2.0154 $\pm$ 0.080 & 2.1457 $\pm$ 0.072 & 2.1598 $\pm$ 0.117 & 2.3557 $\pm$ 0.146 & 2.2736 \\
Max Drawdown & -0.0667 $\pm$ 0.012 & -0.0956 $\pm$ 0.009 & -0.0776 $\pm$ 0.010 & -0.0683 $\pm$ 0.008 & -0.0781 $\pm$ 0.009 & -0.0521 \\
\hline
\end{tabular}
\end{subtable}

\vspace{0.5cm}

\begin{subtable}{\textwidth}
\centering
\caption{Year 2022 (bearish market)}
\begin{tabular}{l|cccccc}
\hline
\textbf{Metric} & \textbf{LLaMA P} & \textbf{LLaMA I} & \textbf{LLaMA I-RAG} & \textbf{LLaMA I-RAG-DF} & \textbf{LLaMA I-RAG-RL} & \textbf{S\&P500} \\
\hline
Cumulative Return & -0.1238 $\pm$ 0.009 & -0.0828 $\pm$ 0.003 & -0.0627 $\pm$ 0.005 & -0.0403 $\pm$ 0.005 & -0.0337 $\pm$ 0.004 & -0.1995 \\
Annualized Return & -0.1251 $\pm$ 0.009 & -0.0837 $\pm$ 0.003 & -0.0635 $\pm$ 0.005 & -0.0407 $\pm$ 0.004 & -0.0341 $\pm$ 0.005 & -0.2016 \\
Annualized Volatility & 0.1471 $\pm$ 0.008 & 0.0990 $\pm$ 0.009 & 0.0788 $\pm$ 0.007 & 0.0780 $\pm$ 0.010 & 0.0608 $\pm$ 0.011 & 0.2416 \\
Sharpe Ratio & -0.8507 $\pm$ 0.077 & -0.8457 $\pm$ 0.083 & -0.8053 $\pm$ 0.096 & -0.5222 $\pm$ 0.084 & -0.5614 $\pm$ 0.131 & -0.8344 \\
Max Drawdown & -0.1590 $\pm$ 0.014 & -0.1037 $\pm$ 0.015 & -0.0967 $\pm$ 0.012 & -0.0901 $\pm$ 0.012 & -0.0640 $\pm$ 0.011 & -0.2543 \\
\hline
\end{tabular}
\end{subtable}
\end{table*}

\subsubsection{Portfolio Construction Based on LLMs Sentiments}

To address \textbf{RQ 3}, we investigated the effectiveness of Large Language Models' (LLMs) sentiment outputs in predicting stock returns. We developed a strategy that involved constructing long-short portfolios and performing backtests. Specifically, trades were executed at the opening price at 9:30 AM on day $T$, with the assumption that all trades were completed at this time. Sentiment scores were derived daily by averaging the sentiment values of ten randomly selected news headlines and tweets, collected between 9:30 AM of the previous day ($T-1$) and 9:30 AM of the current day ($T$). Positive, neutral, and negative sentiments were quantified as 1, 0, and -1, respectively\footnote{For models incorporating RAG, to prevent redundant information, we retrieve information randomly without replacement.}.

We constructed an equal-weighted, zero-cost portfolio that longed (bought) stocks with positive sentiment scores (ranging from 0.1 to 1) and shorted (selled) those with negative sentiment scores (ranging from -1 to -0.1). Scores within the -0.1 to 0.1 range were deemed neutral, prompting us to hold positions. Our analysis was confined to constituents of the S\&P 500 to ensure sufficient liquidity and used this index as a benchmark for U.S. market performance. The analysis covered two periods with different market conditions: 01/01/2021, to 12/31/2021, which provided a scenario of strong recovery and bull market trends—ideal for assessing how the strategy captures growth during economic rebound and low interest rates; and 01/01/2022, to 12/31/2022, which presented a challenging backdrop with high inflation, interest rate hikes, and bear market conditions, testing the strategy's resilience and adaptability to adverse shifts. Transaction costs were set at 10 basis points per trade to cover exchange fees, execution fees, and SEC fees. We tracked the daily cumulative returns of the portfolios constructed using our financial LLMs and compared them with the S\&P 500 as a baseline, as illustrated in \textbf{Figure \ref{fig:model_portfolio}}. Detailed evaluation metrics are shown in \textbf{Table \ref{table:performance_comparison}}.

The results, as depicted in \textbf{Figure \ref{fig:port_2021}}, demonstrate an overall bullish market trend, evidenced by a clear upward trajectory for all LLMs and the S\&P 500 baseline. Despite some volatility, the final cumulative returns at year-end show an increase with the use of more sophisticated LLMs. All models outperformed the baseline market portfolio, which solely invests in S\&P 500 constituent stocks without leveraging sentiment analysis. Notably, only our top-performing model, LLaMA I-RAG-RL, achieved a better Sharpe ratio of 2.3557 compared to the S\&P baseline at 2.2736. This outcome may be attributed to the higher volatility associated with LLM portfolios, suggesting that a more complex trading strategy, such as reinforcement learning, could be advantageous compared to our simpler strategy here that solely relies on stock sentiments.

For the year 2022, as illustrated in \textbf{Figure \ref{fig:port_2022}}, a similar performance enhancement is observed with the adoption of more sophisticated models. Although all LLMs' portfolios registered negative cumulative returns in response to the stock market downturn and overall market decline, these portfolios still managed to mitigate losses and significantly decrease the maximum drawdown compared to the S\&P 500 baseline. The LLaMA I-RAG-DF model achieved a Sharpe ratio comparable to that of the LLaMA I-RAG-DF model but with a slightly larger drawdown, underscoring the effectiveness of this simpler approach. The improved cumulative returns, under adverse market conditions, highlight the capability of our finely tuned and optimized LLMs to generate sentiment outputs that are not only more accurate but also yield more profitable trading signals.

\section{Conclusion and Future Work}
\label{sec:conclusions}

In conclusion, this paper introduces a novel, adaptive retrieval-augmented LLMs framework integrating market feedback for dynamic weight allocation across different knowledge sources within the RAG module. We aim for our financial LLMs to not only extract the real underlying sentiment but also to produce sentiment outputs predictive of stock movements. This is why we choose to design a long feedback loop that refines the weights of the RAG module within the LLMs. Such adjustments affect the context analyzed by the LLMs, influencing the generated sentiment scores. These scores are then assessed against market feedback to further refine the model. Our extensive analysis of benchmark datasets demonstrates that aligning LLMs with market feedback—indirectly through the refinement of RAG weights—enhances both accuracy and F1 scores. The direct refinement approach, while straightforward, is notably effective, especially when employing RL-based methods. Further analysis on portfolio construction shows that LLMs' sentiment outputs can predict the next day's stock price movements. Both analyses indicate that our LLMs' sentiment outputs not only more accurately capture underlying sentiments but are also predictive of stock price movements.

We chose not to employ RL methods like reinforcement learning from human feedback (RLHF) \cite{ouyang2022training} to guide the sentiment output of LLMs directly because it is challenging to design a reward function that accounts for the myriad factors influencing the whole stock market. It is also difficult for humans to ascertain whether a positive sentiment necessarily correlates with a positive market outcome even in the near term. Relying directly on market return signs as rewards might inadvertently focus LLMs on merely predicting price change directions based on textual data, overlooking the extraction of genuine sentiment within the text. Future research should consider the development of a suitable reward function, the training of an effective reward model (RM), and the application of RL methods for direct engagement and enhancement of financial LLMs.

A current limitation of our methodology is its dependence on keyword similarity for information retrieval, risking matches with documents that seem similar but are not semantically related. Future efforts could include developing a semantic search capability within our RAG module by generating vector representations of data for fast retrieval from vector databases. Although keyword searches are simpler, semantic search requires more complex engineering but promises a more effective way for LLMs to access pertinent information.


\bibliographystyle{ACM-Reference-Format}
\bibliography{references}

\FloatBarrier 
\newpage
\appendix

\section{Instruction Set}
\label{sec:instruction}

\section{Detailed workflow example}
\label{sec:details}

\section{State Construction of RL-based Weights Refinement}
\label{sec:state}

\begin{table}[h] 
\centering
\label{tab:rl_state_components}
\begin{tabular}{|p{1.5cm}|p{3cm}|p{3cm}|}
\hline
\textbf{Component} & \textbf{Description} & \textbf{Purpose} \\
\hline
Current Weights & Weights assigned to various knowledge sources within the RAG module. & Indicates the current influence of each source in sentiment analysis. \\
\hline
Historical Accuracy Rate & The accuracy of sentiment predictions over a historical period, measured as a percentage. & Reflects the effectiveness of past predictions in aligning with market movements. \\
\hline
Average Overlap Coefficient & Measures the average similarity between the queries and the retrieved documents, calculated using metrics like cosine similarity. & Assesses the relevance of retrieved information to the query. \\
\hline
Historical Stock Return & The return generated by the stock portfolio over a specified historical period. & Provides insight into the past financial performance. \\
\hline
Technical Indicators & Includes indicators such as MACD, RSI, etc., used to evaluate market conditions. & Helps in identifying trends and potential market turnarounds. \\
\hline
\end{tabular}
\caption{Components of RL State}
\end{table}

\begin{figure}[h!]
\centering
\includegraphics[width=0.48\textwidth]{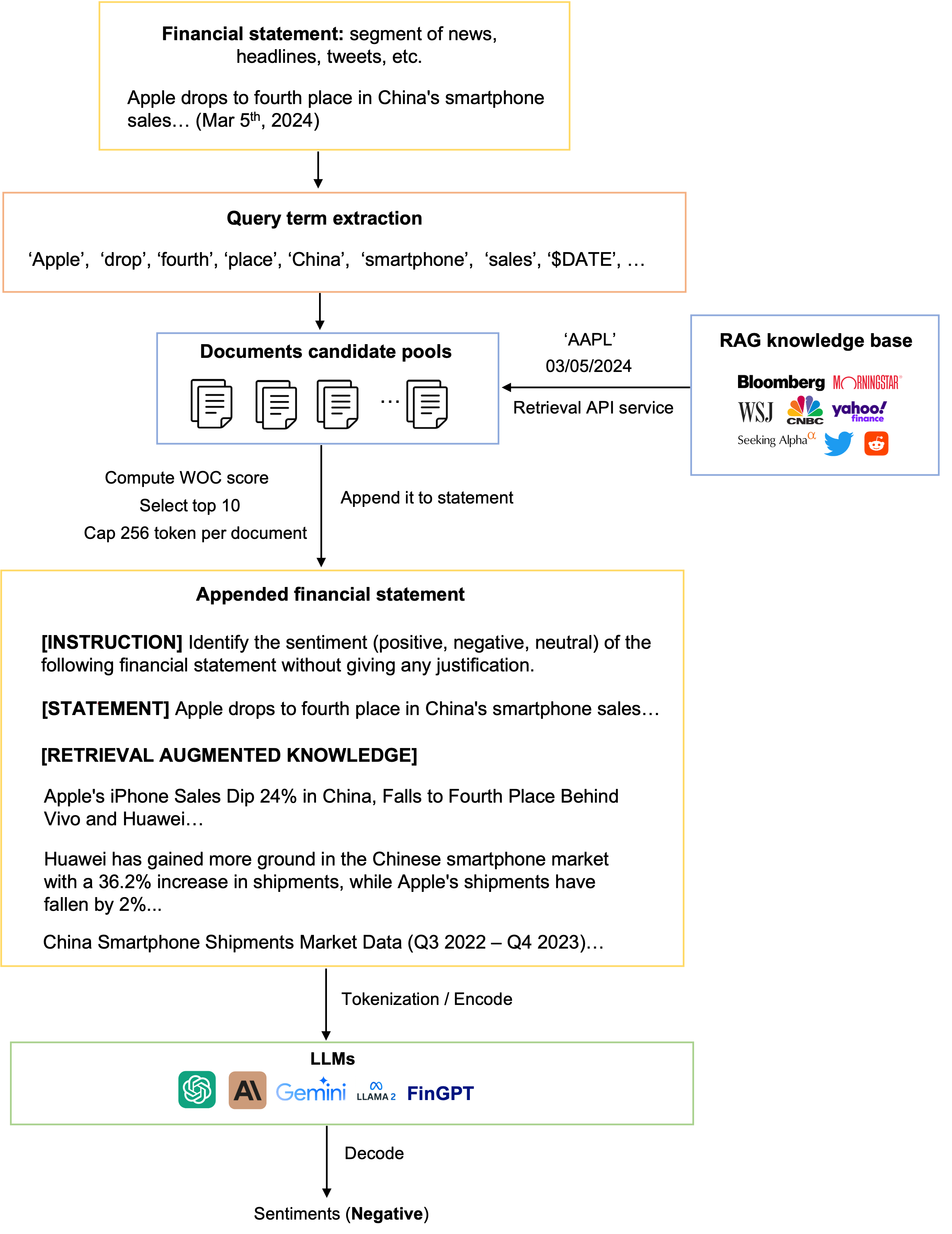}
\caption{Detailed illustrative example for our proposed workflow.}
\label{fig:detailed_workflow}
\end{figure}

\begin{table}[ht!] 
\centering
\begin{tabular}{|c|p{8cm}|}
\hline
\textbf{\#} & \textbf{Instruction} \\ \hline
1 & Identify the sentiment (positive, negative, neutral) of the following financial statement without giving any justification. \\ \hline
2 & Choose the appropriate sentiment: positive, negative, or neutral for the following financial news, and respond only with the sentiment category. \\ \hline
3 & From the options positive, negative, or neutral, select the one that best fits the sentiment of the provided financial news. \\ \hline
4 & Read the following financial statement and classify its sentiment as either positive, negative, or neutral. Provide only the classification. \\ \hline
5 & Determine the sentiment of the following financial news—positive, negative, or neutral—and state your choice clearly. \\ \hline
6 & Evaluate the sentiment of this financial statement as positive, negative, or neutral. Reply with your sentiment classification only. \\ \hline
7 & Assign a sentiment category (positive, negative, or neutral) to the following financial news. Do not include any explanations. \\ \hline
8 & For the financial statement below, indicate whether the sentiment is positive, negative, or neutral. Just provide the sentiment. \\ \hline
9 & Categorize the sentiment of the provided financial news as positive, negative, or neutral. Only the sentiment category is required. \\ \hline
10 & Based on the following financial news or statement, classify its sentiment as positive, negative, or neutral. Simply select your response from the options: positive, negative, or neutral, without providing any explanation. \\ \hline
\end{tabular}
\caption{Instructions for Classifying Financial News Sentiment}
\label{tab:sentiment_instructions}
\end{table}

\end{document}